\documentclass[aps,amsmath,amssymb,groupedaddress,twocolumn]{revtex4}
\usepackage{graphicx}
\pdfoutput=1

\newcommand{\nc}{\newcommand}
\nc{\bs}{\bigskip}
\nc{\beq}{\begin{equation}}
\nc{\eeq}{\end{equation}}
\nc{\beqa}{\begin{eqnarray}}
\nc{\eeqa}{\end{eqnarray}}

\def\gsim{\mathrel{\rlap{\lower4pt\hbox{\hskip1pt$\sim$}}
    \raise1pt\hbox{$>$}}}       

\begin{document}

\title{Physical consequences of the QED theta angle}

\author{Stephen~D.~H.~Hsu} \email{hsu@uoregon.edu}\affiliation{  $~$  \\ 
Academia Sinica, Taiwan \\ and \\ Institute of Theoretical Science, University of Oregon, Eugene, OR 97403 }

\date{December 2010}

\begin{abstract}
We describe a simple gedanken experiment which illustrates the physical effects of the QED theta angle, a fundamental parameter of Nature that has yet to be measured. The effects are manifest in quantum phases analogous to those in the Aharonov-Bohm effect, although they are not intrinsically topological. We also derive the quantum phases using a functional Schrodinger approach, and generalize the results to non-Abelian gauge theories.
\end{abstract}


\maketitle


\bigskip
The gauge symmetry of quantum electrodynamics (QED) allows the introduction in the Lagrangian of a (parity and CP odd) theta term
\beq
{\theta \over 4} \, F_{\mu \nu} \tilde{F}^{\mu \nu} = 
{\theta \over 8} F_{\mu \nu} F_{\rho \sigma} \epsilon^{\mu \nu \rho \sigma} = -  \theta E \cdot B ~.
\eeq
Because this term can be written as a total divergence, it does not alter the classical equations of motion for electromagnetism. Further, it has no influence on amplitudes obtained via perturbation theory -- i.e., ordinary Feynman diagrams. This suggests that the effects of the theta term are non-perturbative and exponentially small -- perhaps of order $\exp( - 1 / \alpha)$, where $\alpha$ is the fine structure constant. 

\begin{figure}[ht]
\includegraphics[width=8cm]{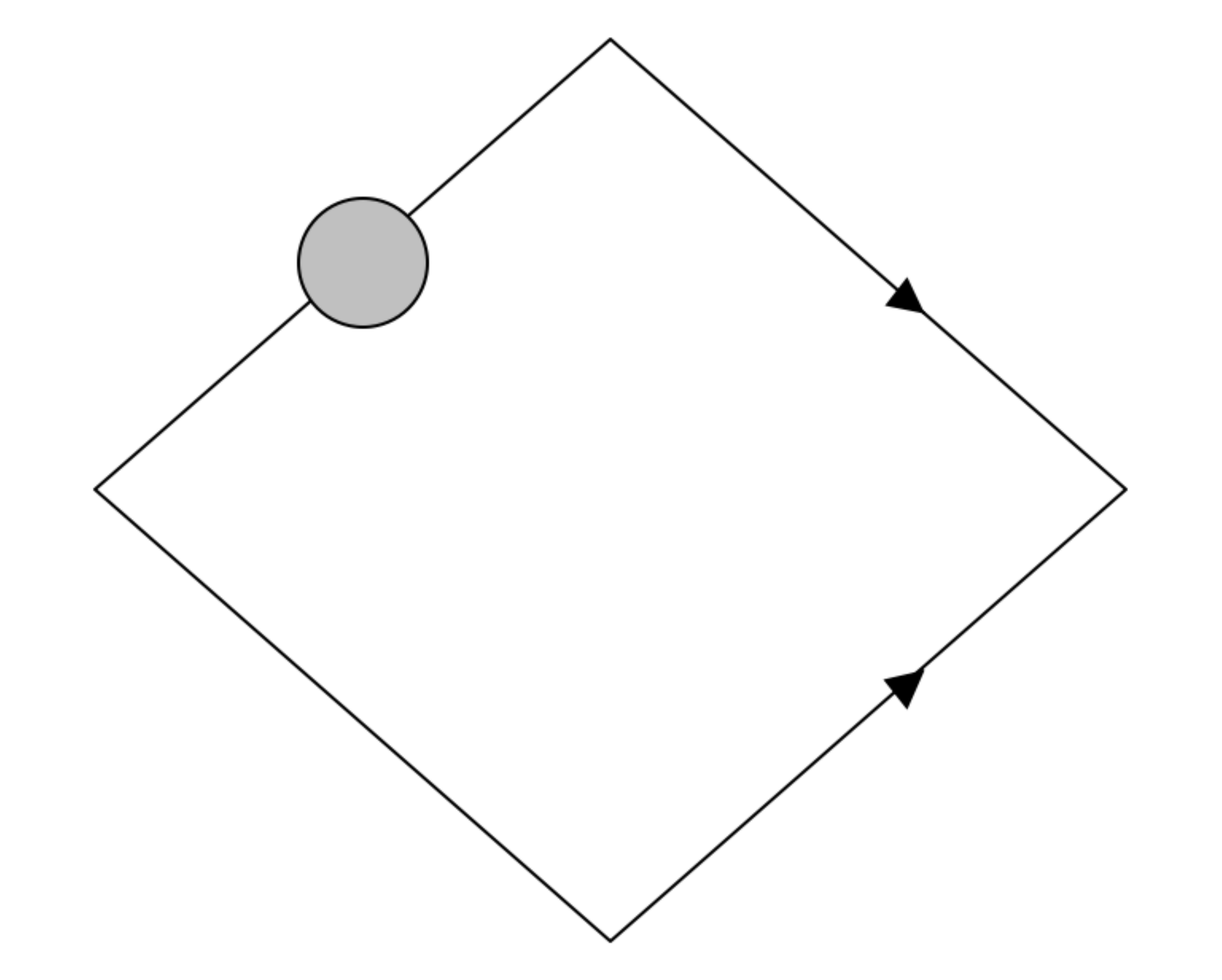}
\caption{Two identical wave packets of light are sent along upper and lower paths of the same length. The upper packet is exposed to a background electromagnetic field depicted by the shaded circle. This background field is chosen so that $E \cdot B$ is non-zero in the interaction region. Interference between the recombined packets depends on the parameter $\theta$.}
\label{figure1}
\end{figure}

However, we shall argue below that the theta term can have significant effects in the presence of strong electromagnetic fields. Consider the gedanken experiment in Figure 1. Two identical wave packets of light are sent along different paths of equal length. (This could be accomplished using a beam splitter and two mirrors or two slits.) The upper packet is exposed to a background electromagnetic field depicted by the shaded circle. This background field is chosen so that $E \cdot B$ is non-zero in the shaded region. $E$ and $B$ represent the {\it combined} fields of the background and wave packet; as a specific example we can take the packet to be polarized with $B$ field perpendicular to the plane of the diagram, and the background field to be an $E$ field along the same direction. The theta term then leads to a relative phase shift between the packets given by
\begin{equation}
\label{phase}
-  \theta  \int d^4x \, E \cdot B ~ \equiv ~ \theta \, \Phi~,
\end{equation}
where the spacetime integral is taken over the region where the packet and background field overlap. Because physical photons have transverse polarization, the theta term is zero everywhere except in the interaction region. The interference pattern obtained when the two packets are recombined depends on $\theta$, which can be measured by varying $\Phi$. The particular arrangement described above is chosen for conceptual simplicity -- it is not meant to be realistic. Indeed it may not be possible to have $E \cdot B$ exactly zero for one path and non-zero for the other. To obtain interference all that matters is that $\Phi$ is not the same for the two different paths \cite{realistic}.

We calculate the phase shift below by computing the quantum amplitude for each wave packet, using the path integral. The combined amplitude is given by
\begin{equation}
{\cal A} (i \rightarrow f) = {\cal A}_{\rm upper} (i \rightarrow f)  + {\cal A}_{\rm lower} (i \rightarrow f)~~. 
\end{equation}
We assume that the packets are described by photon states which are identical except for their directions of propagation, which are related via reflection through the axis of symmetry in Figure 1. We neglect virtual effects from charged particles such as electrons.

Each amplitude can be written (we suppress the index of the vector potential $A_\mu$)
\begin{eqnarray}
\label{pathintegral}
{\cal A} &=& \int dA_i \, dA_f \, \int_{A_i}^{A_f} DA~  \exp \left( - i \int d^4x ~  {1 \over 4} F^2  +   \theta  E \cdot B  \right) \nonumber  \\
&& \Psi^* [A_i] \, \Psi [A_f]~  ~.
\end{eqnarray}
Here $i,f$ denote initial and final times and the wavefunction factors $\Psi$ are the overlap between eigenstates of the field operator $\hat{A}_{\mu} (x)$ and the state describing the packet:
$\Psi [A] = \langle A \vert \Psi \rangle$. In \cite{GHP} it is shown that if this amplitude is evaluated in the stationary phase (i.e., semi-classical) approximation, the $\Psi$ factors determine the appropriate boundary conditions on the classical solution $\bar{A}$ that extremizes the action. The form of  
$\Psi$ and associated boundary conditions for specific cases like plane waves or coherent states are also derived. The overlap factors are assumed to be identical for both packets, up to reflection through the axis of symmetry in the diagram. 

Assuming the photon occupation numbers of the packets and the background field in the interaction region are large, the path integrals can be evaluated in the semi-classical approximation:
\beq
A(x) = \bar{A} (x)  +  \delta A (x)~~,
\eeq
where $\bar{A} (x)$ is the classical field and $\delta A$ is a fluctuation. In the amplitude ${\cal A}_{\rm lower}$ we expand around the solution describing the packet on the lower path, which does not intersect the interaction region, and for which $E \cdot B$ is always zero. In the amplitude ${\cal A}_{\rm upper}$ we expand about the solution $\bar{A} (x)$ describing a packet which propagates through the background field region. There the theta term produces an additional phase factor (\ref{phase}):
\begin{equation}
{\cal A}_{\rm upper} \approx {\cal A}_{\rm lower}~ e^{i \theta \Phi}~,
\end{equation}
where $\Phi$ is understood to be evaluated on the classical field $\bar{A}$, which includes both the wave packet solution and the background field. 

Because the theta term is a total divergence the boundary conditions $A(t_i, x)$ and $A(t_f, x)$ for configurations contributing to ${\cal A}_{\rm upper}$  (with non-zero spacetime integral over $E \cdot B$) are different from the boundary conditions for those contributing to ${\cal A}_{\rm lower}$ (with zero integral over $E \cdot B$). However, in terms of gauge invariant $E$ and $B$ fields both configurations at $t_f$ describe a packet of light incident on the interference point. We stress that physical states are identified by gauge invariant properties; the same physical state has many redundant descriptions due to gauge symmetry. Hence interference of the two amplitudes is still possible \cite{overlap}. 

The integral over fluctuations can be performed exactly because the action remains quadratic even after inclusion of the theta term. Indeed the semi-classical approximation is not necessary, as we now discuss. Since $F_{\mu \nu} F_{\rho \sigma} \epsilon^{\mu \nu \rho \sigma}$ can be written as a total divergence, its contribution to the action can be written as a surface integral over the boundary at, e.g., infinity. The change in the theta term part of the action with respect to local variations of the vector field $\delta A$ is therefore zero as long as the variations do not alter the field values on the boundary. From this it is clear that the equations of motion (deduced from $\delta S[A] / \delta A = 0$) and the second order kernel for fluctuations $\delta^2 S[A] / \delta A^2$ are both independent of $\theta$. We can expand the action $S[A]$ in the path integral about the classical solution $\bar{A}$ (we suppress spacetime integrals):
\begin{equation}
S[A] = S[\bar{A}] + \delta S[A] / \delta A \vert_{\bar{A}} \delta A + \delta^2 S[A] / \delta A^2 \vert_{\bar{A}}\delta A^2 ~.
\end{equation}
The second term on the right vanishes when evaluated on $\bar{A}$ (the classical solution), and the expansion terminates because there are no nonlinear terms. We have noted that $\bar{A}$ and the first and second variations of $S[A]$ are independent of $\theta$. Thus, only the leading term $S[\bar{A}]$ depends on $\theta$, and the dependence has the form given in (\ref{phase}). If we now perform the path integrals exactly for the upper and lower path (including the external field), we will find that even the determinants arising from the integral over fluctuations are equal (they are independent of the theta term), so that the relation
\begin{equation}
{\cal A}_{\rm upper} = {\cal A}_{\rm lower}~ e^{i \theta \Phi}~
\end{equation}
holds exactly.

If the extent of the wave packets is $\sim L$, the size of $\Phi$ can be of order $L^4 \, E \cdot B$. (Because the sign of the $E$ and $B$ components of individual electromagnetic waves oscillates, it is difficult to accumulate phase by increasing the size of the interaction region.) In principle, a sufficiently large background field $E$ could produce a large phase shift for an ordinary laser or microwave pulse with $B$ field aligned along the background field.

We can also deduce our effect using the functional Schrodinger equation for gauge fields, in analogy with the Aharonov-Bohm effect. In the presence of a vector potential the momentum operator for a charged particle becomes $-i \partial_i + A_i$. This leads to the wavefunction phase factor
\begin{equation}
\exp \left( i \int d x \cdot A \right) \psi ~=~ \exp \left( i \int dt \, 
{d x \over dt} \cdot A \right)  \psi
\end{equation}
where $t$ is an affine parameter, such as the time coordinate. A similar result holds in gauge field theory in the presence of the theta term. Recall ${\cal L} = - F^2/4 = 1/2 \left( E^2 - B^2 \right)$. In $A_0 = 0$ gauge $E = -\dot{A}$, so the conjugate momentum $\partial {\cal L} / \partial \dot{A} = \dot{A} = -E$. With no theta term the functional Schrodinger equation  is
\begin{equation}
\frac{1}{2} \left( (-i \delta / \delta A(x))^2 + B(x)^2 \right) \Psi [A]  =  i {\partial \over \partial t} \Psi [A]. 
\end{equation}
The theta term $- \theta \, E \cdot B = \theta \, \dot{A} B$ shifts the conjugate momentum by $\theta B(x)$. Therefore, the  momentum operator $- i \delta / \delta A(x)$ in the Schrodinger equation becomes
\begin{equation}
\label{shift}
-i {\delta \over \delta A(x)} + \theta B(x)~.
\end{equation}
This causes the wave functional $\Psi [A]$ to acquire a phase (analogous to the Aharonov-Bohm phase) associated with motion in the configuration space:
\begin{equation}
\label{Psiphase}
\Psi_\theta [A] ~=~ \exp \left(   i \theta \int d^3x \, {A \cdot B \over 2} \right) \Psi [A] ~.
\end{equation}
The integral is over the timelike component of the current $K^\mu = 1/4 ~ \epsilon^{\mu \alpha \beta \gamma} F_{\alpha \beta} A_\gamma$, which satisfies $\partial_\mu K^\mu = 1/4 \, F \tilde{F} = - E \cdot B$. One can verify that the functional derivative $\delta / \delta A(x)$ of the integral in (\ref{Psiphase}) yields $- \theta B(x)$, which cancels the shift in (\ref{shift}). Thus, for $\Psi [A]$ a solution of the Schrodinger equation in the absence of a theta term, $\Psi_\theta [A]$ is the corresponding solution when the theta term is added to the Lagrangian. We can define the phase relative to that of a reference configuration $A_* (x)$:
\begin{equation}
\label{motion}
 i \theta \left( \int d^3x \, K^0 (A) -   \int d^3x \, K^0 (A_*) \right) ~=~ 
 - i  \theta \int d^4x \, E \cdot B~,
\end{equation}
where appropriate boundary conditions are imposed on the spacetime integral (see below). Once $A^*$ is fixed the phase factor is determined for all configurations $A$ and for all times. Motion in configuration space, analogous to motion in coordinate space for the Aharonov-Bohm case, means a trajectory from the reference configuration $A_* (x)$ to the configuration of interest $A(x)$.

The meaning of Equation (\ref{motion}) is as follows. In the presence of the theta term, the Schrodinger wave functionals $\Psi_\theta [A]$ evaluated on two different configurations $A_i(x)$ and $A_f(x)$ acquire an additional phase relative to each other given by the rhs of (\ref{motion}), where the integral is taken over a path in configuration space with boundary conditions $A(t_i, x) = A_i(x)$ and $A(t_f,x) = A_f(x)$, and the limits of $t$ integration are $t_i$ and $t_f$ (here $t$ is an affine parameter, not necessarily time). The total divergence property of $E \cdot B$ guarantees that any choice of interpolation yields the same phase for given $A_i(x)$ and $A_f(x)$, since the integrals only depend on the boundary conditions. For the following discussion, it is useful to define the phase factor given above as 
$i \theta \Phi [ A_f \vert A_i ]$. Note that $\Phi [ A_1 \vert A_3 ] = \Phi [ A_1 \vert A_2 ] + \Phi [ A_2 \vert A_3 ]$.

We now apply these results to our gedanken experiment, with $A$ representing the entire gauge field configuration (both wave packet and background field). We wish to show that, in the presence of the theta term, the phase shift $\Phi [ A_{f, {\rm upper}} \vert A_{f, {\rm lower}} ]$ is as defined in (\ref{phase}), or, equivalently,
\begin{equation}
\Phi [ A_{f, {\rm upper}} \vert A_{f, {\rm lower}} ] = \Phi [ A_{f, {\rm upper}} \vert A_{i, {\rm upper}} ]~.
\end{equation}
Here $A_{f, {\rm upper}}$ is the final configuration for the case where the packet follows the upper path and intersects the background field, $A_{f, {\rm lower}}$ is the corresponding final configuration for the lower path, and $i$ denotes initial rather than final. First we write
\beqa
&& \Phi [ A_{f, {\rm upper}} \vert A_{f, {\rm lower}} ] = \Phi [ A_{f, {\rm upper}} \vert A_{i, {\rm upper}} ] + \nonumber \\
&&  ~~~\Phi [ A_{i, {\rm upper}} \vert A_{i, {\rm lower}} ]
+  \Phi [ A_{i, {\rm lower}} \vert A_{f, {\rm lower}} ]~.
\eeqa
Next, we observe that $\Phi [ A_{i, {\rm upper}} \vert A_{i, {\rm lower}} ]$ is zero by assumption -- the two initial states are produced with no relative phase, e.g., by a perfect beam splitter. (Alternatively, one can compute this phase using an interpolation to find that it is zero because $E \cdot B$ is always zero; the initial states for the wave packets are far from the background field region.) Finally, $ \Phi [ A_{i, {\rm lower}} \vert A_{f, {\rm lower}} ]$ is zero because the interpolation between the initial and final configurations on the lower trajectory have $E \cdot B = 0$ at all times. A crucial assumption in our construction is that one can, at least in principle, arrange for the spacetime integral over $E \cdot B$ to be different for each path in configuration space. The arrangement described in the gedanken experiment is simply one example \cite{realistic}.

The same functional Schrodinger calculation can be repeated in non-Abelian gauge theories, with the resulting phase determined by the topological charge density ${\rm tr} F \tilde{F}$ or corresponding current
\begin{equation}
K^\mu = \epsilon^{\mu \alpha \beta \gamma} {\rm tr} \left( F_{\alpha \beta} A_\gamma - {2 \over 3} A_\alpha A_\beta A_\gamma \right)~.
\end{equation}
In \cite{Treiman:1986ep} it was shown that the wave functional $\Psi [A^U] = \exp(i \theta \Phi) \Psi[A]$, where $A$ and $A^U$ are related by a gauge transformation and the phase $\Phi$ is defined as above for $A_i = A$ and $A_f = A^U$. For vacuum configurations the phase is only non-zero if the gauge transformation $U(x)$ is topologically nontrivial, and in this case $\Phi$ is quantized. For generic gauge configurations $A_i(x)$ and $A_f(x)$ (i.e., not necessarily vacuum configurations, nor related by a gauge transformation) the topological charge is not quantized, but rather takes on continuous values \cite{GH}. Therefore, quantum phases of the type discussed here can be found in non-Abelian theories even in the absence of nontrivial topology.

\smallskip

As we have seen, the theta term can have a quantum mechanical effect on local physics despite the fact that it is a total divergence and has no effect on the classical equations of motion (cf. the Aharonov-Bohm effect). Although the spacetime integral of $E \cdot B$ over a region is fixed by the values of the potential $A$ on the boundary, the specific arrangement of the density $E \cdot B$ within the region can lead to observable consequences: relative phases for different photon states. This is not so different from the case of the electric charge $Q$: the total $Q$ on a spacelike slice is fixed, but the distribution of charge density has local consequences. Similarly, in non-Abelian gauge theories (e.g., QCD), local fluctuations in topological charge density can have physical effects even if the boundary conditions (and hence total topological charge) are held fixed. 

These effects violate CP symmetry, so it is possible they may have some relevance to the baryon asymmetry of the universe. The SU(2) theta angle has no physical consequences, because it can be canceled by appropriate chiral rotation of the left handed fermions. But electroweak baryon number violating processes (i.e., mediated by sphaleron-like configurations) typically involve strong electromagnetic fields, so might be affected by the CP violating QED theta angle.


In grand unified theories such as SU(5) or SO(10), the theta angles for each of the standard model gauge forces (i.e., SU(3), SU(2), U(1)) are related by group theoretical factors. Therefore, low energy measurements of these angles have interesting implications for very high energy physics.

\bigskip


\emph{Acknowledgments ---} The author thanks D. Reeb, D. Soper, T.C. Yuan and A. Zee for discussions and Academia Sinica for its hospitality. This work is supported by the Department of Energy under grant DE-FG02-96ER40969.

\bigskip



\begin{thebibliography}{99}
\bibitem{realistic} Rather than the spacetime integral over $E \cdot B$ we can consider the boundary integrals over $K^0 = 1/2 \,A \cdot B$. Writing $A(x) = A_b (x) + A_p (x)$ as the sum of background and packet, only the mixed terms $A_p \cdot B_b$ and $A_b \cdot B_p$ contribute to the phase difference. It seems plausible that for many choices of background the boundary integrals at $t_f$ and $t_i$ will give different results for upper and lower packets which move in different directions and therefore have different $A_p$ and $B_p$ fields. Thus $\Phi$ will differ along the upper and lower paths.
\end{thebibliography}
\end{document}